\documentclass[twocolumn,twoside,slac_two]{revtex4}
\usepackage{graphicx}
\usepackage{fancyhdr}
\usepackage{graphicx}% Include figure files

\newcommand{\nl}{\nonumber\\}

\newcommand{\be}{\begin{equation}}
\newcommand{\ee}{\end{equation}}
\newcommand{\bea}{\begin{eqnarray}}
\newcommand{\eea}{\end{eqnarray}}

\begin{document}

\title{Progress in lattice QCD}


\author{Tetsuya Onogi}
\affiliation{Yukawa Institute for Theoretical Physics, Kyoto, Japan}

\begin{abstract}
I review the recent progress in lattice QCD, which will be useful in
heavy quark physics in the near future. Reviewing the theoretical
developments in lattice QCD first, I focus our recent unquenched QCD
with dynamical overlap fermion as implemented by JLQCD collaboration. I also
introduce some of our recent studies on the $B^*B\pi$ coupling and on
the determination of $|V_{ub}|$ through the dispersive bound. 
\end{abstract}

\maketitle

\thispagestyle{fancy}


%%%%%%%%%%%%%%%%%%%%%%%%%%%%%%%%
\section{Introduction}

The lattice computation of weak matrix elements can be defined as 
\begin{equation}
\langle {\cal O}^{cont}(\mu) \rangle
\equiv 
\lim_{a\rightarrow 0,
m_i \rightarrow m_i^{\rm phys}}
Z(a\mu,g_0(a)^2) \langle {\cal O}^{lat}(a) \rangle,
\end{equation}
where
\begin{equation}
\langle {\cal O}^{lat}(a) \rangle 
\equiv 
\frac{\int [dU] \prod_{i=u,d,s} det[D(m_i)] e^{-S_g(U)}
{\cal O}^{lat}(a)}
{\int [dU] \prod_{i=u,d,s} det[D(m_i)] e^{-S_g(U)}}.  
\end{equation}
In order to extract physical matrix elements from the lattice 
the sea quark effects, the renormalization factor, 
and the continuum limit and chiral limit should be incorporated.
In addition, the heavy quark should be treated either by the 
effective theory or by extrapolation in the heavy quark mass 
from the smaller mass regime. These steps have been quite nontrivial 
tasks. Recently, there are three major progresses 
in lattice QCD which can drastically reduce the systematic errors in
lattice QCD computation: (1) unquenched lattice QCD simulations in the
chiral 
regime, (2) nonperturbative renormalization, and (3) a new approach 
to heavy quarks on the lattice. In the following I explain these
developments in some detail. 
\subsection{Unquenched QCD simulations in the chiral regime}
A few years ago the only available large scale dynamical QCD
simulations with lightest pion mass $m_{\pi}\simeq 300$ MeV, 
was by the MILC collaboration with the staggered quark~\cite{milc07}. This
is  because of the smallest numerical cost owing to the small degrees 
of freedom of the fermion and numerical stability from the exact 
partial chiral symmetry. In contrast, it has been thought that
light dynamical fermion simulations in other fermion formalism 
would be difficult by Tflops machines. 

In recent years, O(10) Tflops machines have become available in many places.
Also the new preconditioners for Dirac operator inversion algorithm 
such as 'domain decomposition'~\cite{luscher} and 'mass
preconditioning'~\cite{massprec} enables us to treat the
high- and low-mode contributions  to the Dirac operator
separately. Combining this method with the  
'multi-time scale' in the molecular dynamics step~\cite{sexton} 
gives significantly efficient algorithms for updating the gauge
configurations in the hybrid Monte-Calro method. 
Owing to theses developments, there are now may unquenched simulations
~\cite{del06,Aoki:2008sm,urbach07,rbcukqcd08,Noaki:2008iy} 
as shown in Table~\ref{tab:simulation}. 
\begin{table*}[t]
\begin{center}
\begin{tabular}{l|l|l|l|l|l}
Group & Fermion Action & $n_f$ & $a$(fm)& $L$(fm) & $m_{\pi}$(GeV)\\
\hline
MILC~\cite{milc07}  & Improved Staggered & 2+1 & 0.09, 0.12 & 3 & $\geq 300$\\
CERN~\cite{del06}
  & Wilson, O(a)-imp Wilson & 2 & 0.052-0.075 & 3 & $\geq 300$\\
PACS-CS~\cite{Aoki:2008sm}
 & O(a)-imp Wilson & 2+1 & 0.07, 0.10, 0.12 & 3 & $\geq 210$\\
ETMC~\cite{urbach07}
  & twisted mass Wilson & 2 & 0.075, 0.096 & 3 & $\geq 270$\\
\hline
RBC/UKQCD~\cite{rbcukqcd08}
  & Domain-wall & 2+1 & 0.12 & 3 & $\geq 330$\\
JLQCD~\cite{Noaki:2008iy}
  & Overlap & 2,2+1 & 0.12 & 2 & $\geq 300$\\
\end{tabular}
\label{tab:simulation}. 
\caption{Projects of large scale lattice QCD simulations with light
dynamical quarks.}
\end{center}
\end{table*}
Although the staggered fermion simulations is going much ahead,
new results in other approaches  will give important 
numerical and theoretical cross-checks with different advantages 
and disadvantages. 
\subsection{Nonperturbative renormalization}
Another major development is the proposal of renormalization schemes
with which lattice simulation can give the renormalization factors
nonperturbatively.  

One such scheme is the Schrodinger functional (SF-) scheme which is
defined by amplitudes in a finite box with physical size $L$ with
Dirchlet boundary conditions in the temporal direction~\cite{luscher92}
. The physical amplitudes to define this scheme are computable both
perturbatively and nonperturbatively. The renormalization scale is
defined as $\mu=1/L$. In this scheme both the renormalization constant
of any local operators and their running can be obtained. To obtain
the running, additional simulations with different box sizes are
needed. 

Another scheme is the regularization independent momentum scheme
(RI-MOM) defined by off-shell quark/gluon amplitudes in Landau
gauge~\cite{Martinelli:1994ty}. The amplitudes are also computable
both perturbatively and 
nonperturbatively. The renormalization scale is defined by the
momentum scale.   
\subsection{New approach to the heavy quark}
The precise computation of weak matrix elements of the B meson 
is one of the most important topics in lattice QCD. However 
since the typical lattice cutoff used in practical simulations is
smaller than the bottom quark mass, naive lattice methods suffer from 
a large discretization error. For this reason lattice nonrelativistic
QCD action has been widely used. Unfortunately, due to the
nonrenormalizability of the action, one cannot take the continuum
limit in this approach so that one suffers from sizable systematic 
errors from discretization error and perturbative renormalization error.
Recently new methods to treat the heavy-light meson in the continuum
limit with nonperturbative accuracy have been proposed. 

The first approach was proposed by Alpha
collaboration~\cite{Heitger:2003nj}. They used the lattice HQET
which is matched to QCD with nonperturbative accuracy by Schrodinger 
functional method in small volume with sufficiently fine lattice. 
Then, they evolve the lattice HQET action to coarser lattice by step
scaling. They can also include $1/M_b$ corrections into the action and
operators.

The second approach is the step scaling method proposed by the Rome II
 group~\cite{deDivitiis:2003iy,deDivitiis:2003wy}.
The compute the physical observable (e.g. $f_B$ ) in small volume 
with $L_0=0.4$ fm with sufficiently fine lattice using relativistic
 quark action. Then, they compute the finite size corrections 
with larger volumes $L= 2 L_0,  4 L_0$ by extrapolations from smaller
 heavy quark masses as
\begin{eqnarray}
f_{B_s}(L_{\infty}) = f_{B_s}(L_0)  \sigma( L_0) \sigma( L_1),
\end{eqnarray}
where $\sigma(2L_0) = \frac{f_{B_s}(2L_0)}{f_{B_s}(L_0)}$ and
$\sigma(L_{\infty}) = \frac{f_{B_s}(L_{\infty})}{f_{B_s}(2L_0)}$.
Guazzini et al.~\cite{Guazzini:2007ja} combined the above two methods,
i.e. they use the static limit result to interpolate the finite volume
corrections.

Table~\ref{tab:fB} shows the quenched QCD results of $f_B$ with
nonperturbative accuracy. It is remarkable that all three approaches  give 
consistent results with high accuracy.
\begin{table}[h]
\begin{center}
\begin{tabular}{l|l}
Method & $f_{B_s}$ (MeV)\\
\hline
HQET with $1/M$ (Alpha)~\cite{Rolf:2003mn}& 193(7)\\
Step scaling (Rome II)~\cite{deDivitiis:2003wy} & 195(11)\\
Combination~\cite{Guazzini:2007ja}            & 191(6)
\end{tabular}
\end{center}
\label{tab:fB}
\caption{Quenched QCD results of $f_B$ with nonperturbative accuracy.}
\end{table}
\section{Recent results with dynamical overlap fermion}
The fermion action satisfying the Ginsparg-Wilson
relation~\cite{Ginsparg:1981bj} 

\begin{eqnarray}
D\gamma_5+\gamma_5 D = a D\gamma_5 D
\end{eqnarray}
realizes the exact chiral symmetry on the
lattice~\cite{Luscher:1998pqa} 
\begin{eqnarray}
\delta \psi = \gamma_5 ( 1-aD) \psi,
\delta \bar{\psi} = \bar{\psi}\gamma_5.
\end{eqnarray}
An explicit construction of the Ginsparg-Wilson 
fermion called as the overlap fermion was 
proposed by Neuberger, which is defined as 
\begin{eqnarray}
D= \frac{1}{a}\left[ 1+ \gamma_5 \epsilon({H_W})\right],
\end{eqnarray}
where $H_W=\gamma_5(D_W+M_0)$ is the Wilson Hamiltonian with negative
mass term $M_0$ at the cutoff scale~\cite{overlap}. 
The domain-wall fermion is
another realization of the Ginsparg-Wilson fermion which introduce
5-th dimension~\cite{DW}. However, with finite extent in the 5-th
dimension, the chiral symmetry becomes approximate with an
exponentially suppressed symmetry violation. 

The JLQCD collaboration succeeded in the first large scale lattice QCD
simulation with a dynamical overlap fermion. $N_f=2$ unquenched
simulation on a $16^3\times 32$ lattice at $a=0.12$ fm was carried out
for 6 quark masses covering the range $m_s/6\sim m_s$ and various
measurements of physical observables were
made~\cite{Ohki:2008ff,Shintani:2008qe,Aoki:2008tq,Fukaya:2007pn,Aoki:2007pw,Fukaya:2007yv,Fukaya:2007fb}    
\subsection{Chiral behavior of $m_{\pi}$ and $f_{\pi}$}
\begin{figure*}[tb]
\centering
\includegraphics[width=80mm]{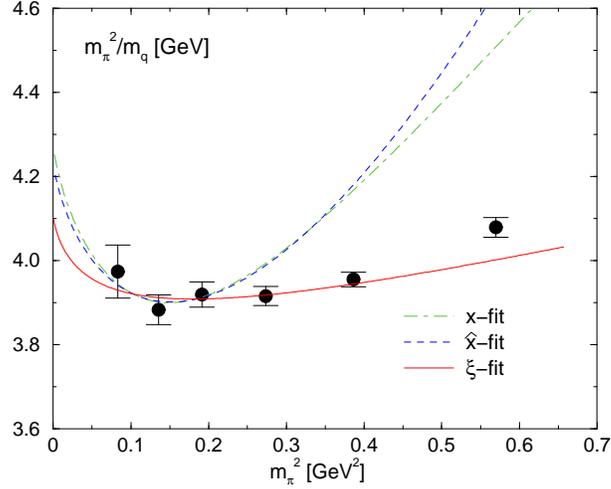}
\caption{ChPT fit with $x$-expansion and $xi$-expansion. 
$x$-expansion makes the convergence of the ChPT fits better.}
\label{fig:mpi_mud_JLQCD}
\end{figure*}

JLQCD collaboration studied the quark mass dependence of  $m_{\pi}$
and $f_{\pi}$ for $n_f=2$ QCD with dynamical overlap
fermion~\cite{Noaki:2008iy}. They found that the lattice data for
$m_{\pi}\leq 450$ MeV are well fitted with NLO ChPT formula
\begin{eqnarray}
m_{\pi}^2/m_q &=& 2 B ( 1 + x \ln x ) + c_3 x\\
f_{\pi} &=& f ( 1  - 2  x \ln x ) + c_4 x,
\end{eqnarray}
with $x\equiv m_{\pi}^2/(4\pi f)^2$ as shown in
Figs.\ref{fig:mpi_mud_JLQCD}. They also studied the convergence of
the ChPT by replacing the expansion parameter $x$ by 
$\hat{x}\equiv 2 m_{\pi}^2/(4\pi f)^2$ or $\xi\equiv m_{\pi}^2/(4\pi
f_{\pi})^2$. They find that the NNLO ChPT with $\xi$-expansion can nicely
describe the lattice data in the pion mass region of $290 \sim 750$
MeV. 
\subsection{$B_K$}
Indirect CP violation in the K meson system $\epsilon_K$ is one of 
the most crucial quantities used to test the standard model and the physics
beyond. The experimental value is determined with high accuracy as
\begin{eqnarray}
|\epsilon_K|= (2.233\pm 0.015)\times 10^{-3}.
\end{eqnarray}
Theoretically this quantity is described as
$|\epsilon_K|=  f(\rho,\eta) \times C(\mu) \times B_K(\mu)$.
Here, $f(\rho,\eta)$ is a factor which depends on the CKM matrix
elements, $C(\mu)$ is the Wilson coefficient from short-distance
QCD corrections and $B_K(\mu)$ is the bag parameter defined as
\begin{eqnarray}
B_K(\mu) = \frac{\langle K^0 | \left[\bar{d}\gamma^{\mu}(1-\gamma_5) s
\bar{d}\gamma_{\mu}(1-\gamma_5) s \right](\mu) | \bar{K}^0
\rangle}{\frac{8}{3} f_K^2 m_K^2}. 
\end{eqnarray}
The main problem in unquenched lattice calculations is the possible
operator mixing in wrong chiralities or tastes. The overlap fermion is
free from operator mixing owing to the exact chiral symmetry.
The JLQCD collaborations study $B_K$ with overlap fermion 
in 2 flavor QCD at lattice spacing $a=0.12$ fm on physical volume with
$L=2fm$~\cite{Aoki:2008ss}. They have 4 points for the sea quark and
10 combinations of valence quark masses  
$(m_1,m_2)$. They fit the data with NLO PQChPT. The renormalization 
factor is determined nonperturbatively by the RI-MOM scheme. They obtain 
$\hat{B} = 0.734(5)_{\rm stat.}(50)_{\rm sys.}$, where the dominant 
error comes from the finite size effect of order  $5\%$.

It should be noted that the long standing operator mixing problem 
is solved with the advent of an overlap fermion with the exact chiral
symmetry. Thus the above study heralds the beggining precision studies 
of $B_K$ for which  significant progress will be expected in near future. 

%%%%%%%%%%%%%%%%%%%%%%%%%%%%%%%%%%%%%%%%%%%%%%%%%%%%%%%%%%%%
\section{Some new results in heavy quark physics}

\subsection{Determination of the $B^* B\pi$ coupling}
The $B^* B\pi$ coupling is a fundamental parameter of chiral 
effective Lagrangian with heavy-light mesons defined as
\begin{equation}
L= -\mbox{Tr}\left[\bar{H} iv \cdot D H\right]
+ \hat{g}_b \mbox{Tr}\left[\bar{H}H A_{\mu}\cdot \gamma_{\mu}\gamma_5 
\right] +O(1/M),
\end{equation}
where the low energy constant $\hat{g}_b$ is the $B^*B\pi$ coupling,
$v$ is the four-velocity of the heavy-light meson $B$ or $B^*$,
and $H$, $D_{\mu}$, $A_{\mu}$ are described by the $B$, $B^*$ and $\pi$
fields as $H =
  \frac{1}{2}(1+\gamma_{\mu}v_{\mu})(iB\gamma_5+B_{\mu}^*\gamma_{\mu})$,
$\xi=\exp(i\pi/f)$, $ D_{\mu} = \partial_{\mu}+\frac{1}{2}
  (\xi^{\dagger}\partial_{\mu}\xi +\xi\partial_{\mu}\xi^{\dagger})$,
$A_{\mu} = \frac{i}{2}
(\xi^{\dagger}\partial_{\mu}\xi -\xi\partial_{\mu}\xi^{\dagger})$.

Once the $B^*B\pi$ coupling is determined, the heavy meson effective
theory can predict various quantities which are important for CKM
phenomenology~\cite{Boyd:1994pa}. 
For example the light quark mass dependence of the $B$ meson decay
constant can be determined as  
\begin{equation}
f_{B_d}= F \left( 1 + \frac{3}{4}(1+3\hat{g}_b^2) \frac{m_{\pi}^2}{(4\pi
f_{\pi})^2} \log(m_{\pi}^2/{\Lambda^2})\right) + \cdots,
\end{equation}
$F$ is the low energy constant associated with the heavy-light 
axial-vector current. 
The form factor $f^+(q^2)$ for the semileptonic decay
$B\rightarrow\pi l\nu$ can also be expressed
in terms of the $B^*$ meson decay constant $f_{B^*}$ and
$\hat{g}_b$ as
\begin{eqnarray}
f^+(q^2)
=-\frac{f_{B^*}}{2f_\pi}
  \left[ \hat{g}_b \left(
      \frac{m_{B^*}}{v \cdot k-\Delta}-\frac{m_{B^*}}{m_B}
  \right) +\frac{f_B}{f_{B^*}} \right],
\end{eqnarray}
where $v$ is the velocity of the $B$ meson,
$k$ is the pion momentum, and $\Delta=m_{B^*}-m_{B}$.
Therefore, the precise determination of the $B^*B\pi$ coupling
is crucial for determining $|V_{ub}|$ and  $|V_{td}|$ accurately.
Despite its importance the $B^*B\pi$ has been known not so
accurately due to the large statistical error of the heavy-light
meson in static limit~\cite{deDivitiis:1998kj,Abada:2003un,Becirevic:2005zu}.

We carry out a precise determination of the $B^* B\pi$ coupling in 
$n_f=2$ QCD  using 100 to 150 gauge
configurations  provided by CP-PACS collaboration~\cite{Ali
Khan:2001tx} through JLDG (Japan Lattice DataGrid), which $12^3\times
24$ lattices  at $\beta=1.80$ and $16^3\times 32$ lattices at
$\beta=1.95$ with two flavors of $O(a)$-improved Wilson quarks 
and the Iwasaki gauge action.
In order to improve the statistical signal, we exploit 
the static quark action using the HYP smeared links~\cite{Hasenfratz:2001hp}
with the smearing parameter values 
$(\alpha_1,\alpha_2,\alpha_3)=(0.75,0.6,0.3)$. We also employ the
all-to-all propagator~\cite{Foley:2005ac} using the low-mode
averaging technique~\cite{DeGrand:2002gm,Giusti:2004yp}. Based on the
previous quenched study~\cite{Negishi:2006sc}, we take 200 low-eigenmodes.

The physical value of the $B^* B\pi$ coupling is obtained by
multiplying the bare value by the renormalization constant at one-loop.
The chiral extrapolation is made in three ways:
 (a) the linear extrapolation,
(b) the quadratic extrapolation, and (c) the quadratic plus chiral
log extrapolation where the log coefficient is determined from ChPT
\cite{Cheng:1993kp}. We take the result at $\beta=1.95$ 
as our best estimate for the physical value of $\hat{g}_\infty$,
and estimate the discretization error of $O((a\Lambda)^2)$
by order counting with $\Lambda \sim 0.3$ GeV.
Including the perturbative error of $O(\alpha^2)$ also by
order counting, our result for $\hat{g}_\infty$ is 
\begin{equation}
\hat{g}_\infty^{n_f=2} 
=  0.516(5)_{\text{stat}}(31)_{\text{chiral}}
        (28)_{\text{pert}}(28)_{\text{disc}} .
\label{eq:final_result}
\end{equation}
\begin{figure}[tb]
\vspace*{0.5cm}
\begin{center}
\includegraphics[width=7.8cm,clip]{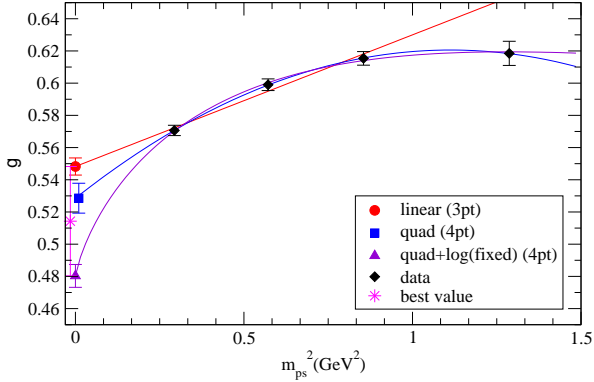}
\caption{
The chiral extrapolation of the physical $B^*B\pi$ coupling 
at $\beta=1.95$.
}\label{fig:chextr12_16}
\end{center}
\end{figure}
%
%%%%%%%%%%%%%%%%%%%%%%%%%%%%%%%%%%%%%%%%%%%%%%%%%%%
\subsection{Dispersive bounds on the form factor for $B\rightarrow\pi
l\nu$ decay}
The momentum range of $B\rightarrow \pi l\nu$ form factors computed 
from lattice QCD is limited by the small recoil or large $q^2$
region. This leads to a big disadvantage because most of the
experimental data lies in large recoil region. While one can
extrapolate in $q^2$ with a fit ansatz, this will always introduce 
some model dependence. Dispersive bounds is one possible way to
constrain the $q^2$ dependence in model independent fashion using
unitarity. 
\begin{figure}[h]
\begin{center}
\includegraphics[width=8cm]{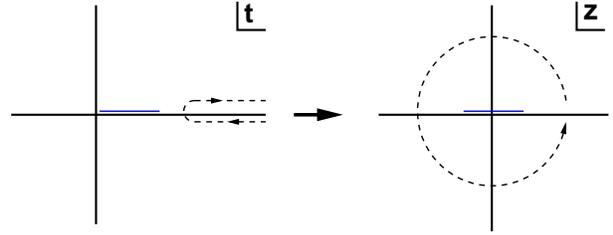}
\caption{A map from $t$ plane to $z$ plane}
\end{center}
\label{fig:map}
\end{figure}
Consider the imaginary part of the vacuum polarization amplitude for
the current $V(x)=\bar{u}\gamma_{\mu}b(x)$ and a map as in 
Fig.~\ref{fig:map}
\begin{eqnarray}\label{eq:ope}
!\!\\Pi^{\mu\nu}(q) 
& \equiv & i  \int d^4x\, e^{iq\cdot x} \langle 0 | T\left\{ V^\mu(x)
V^{\nu\dagger}(0) \right\} | 0 \rangle  \nonumber\\ 
&=& (q^\mu q^\nu - g^{\mu\nu} q^2)\Pi_1(q^2) + q^\mu q^\nu \Pi_0(q^2),
\end{eqnarray}
Then,  from dispersion relations one obtains 
\begin{eqnarray}
\chi_{F_+}(Q^2) &=& {1\over 2}{\partial^2 \over \partial (q^2)^2}
\left[ q^2 \Pi_1 \right]  = {1\over \pi} \int_0^\infty\! dt\, {t {\rm
Im}\Pi_1(t) \over (t+Q^2)^3 } \,, \nl \chi_{F_0}(Q^2) &=& {\partial
\over\partial q^2} \left[ q^2 \Pi_0 \right]  = {1\over \pi}
\int_0^\infty\! dt\, {t {\rm Im}\Pi_0(t) \over (t+Q^2)^2 } \,. 
\label{eq:disp}
\end{eqnarray} 
with $Q^2= -q^2$ and $\eta$ an isospin factor, while $\chi$ 's can be
computed using the OPE and perturbative QCD. 
Unitarity tells us that this is equal to the sum over all the hadronic
states. and dropping all the excited states and leaving only 
$B\pi$ and $B^*$ states gives an exact bound. 
\begin{eqnarray}
{\eta \over 48\pi}{ [(t-t_+)(t-t_-)]^{3/2} \over t^3} |F_+(t)|^2
&\le& {\rm Im} \Pi_1(t) \,, \nl
{\eta t_+ t_- \over 16\pi}{ [(t-t_+)(t-t_-)]^{1/2} \over t^3} |F_0(t)|^2 &\le& 
{\rm Im} \Pi_0(t) \,, 
\label{eq:bound}
\end{eqnarray} 
Combining Eqs.~\ref{eq:disp},~\ref{eq:bound}
and making change of variables in the integration from $t$ to $z$
\begin{equation}
z(t,t_0) = \frac{\sqrt{t_+-t}-\sqrt{t_+-t_0}}
          {\sqrt{t_+-t}+\sqrt{t_+-t_0}},  
\end{equation}
with $t_{\pm}=(m_B\pm m_{\pi})^2$, 
one obtains
\begin{eqnarray}
\langle \phi f_0 | \phi f_0 \rangle < \chi_0, & 
\langle P \phi f_+ | P \phi f_+ \rangle < \chi_+,
\end{eqnarray}
where J is a quantity which can be obtained using OPE and perturbative
QCD. The inner product $\langle g | h \rangle$ for arbitrary functions
$g(z)$ and $h(z)$  is defined by the integral along the unit circle in
the $z$ plane as  
\begin{eqnarray}
\langle g | h \rangle \equiv \int \frac{dz}{2\pi i} (g (z))^{\ast}.
\end{eqnarray}
$P(z)=z(t,m_B^*)$ is multiplied by $f_+$ in order  to remove the $B^*$ pole
inside the unit circle.  Cauchy's theorem tells that if we know the 
additional integrated quantity $\langle g_i | P \phi_+ f_+ \rangle$ 
with a set of known functions $\{g_i(z), i=1,...,N \}$ one can make the
bound stronger as  
\begin{equation}
\det \left( 
\begin{array}{cccc}
 \chi     &\langle P\phi f_+ | g_1 \rangle & \ldots 
       &\langle P\phi f_+ | g_N  \rangle \\
        \langle g_1  | P\phi f_+ \rangle 
       &\langle g_1 | g_1 \rangle & \ldots
       &\langle g_1 | g_N \rangle \\
\vdots &\vdots & \ddots & \vdots\\
        \langle g_N  | P\phi f_+ \rangle 
       &\langle g_N | g_1 \rangle & \ldots
       &\langle g_N | g_N \rangle \\
 \end{array}
\right) > 0 .
\end{equation}
Choosing $g_n(z) = \frac{1}{z-z(t)}$, Lellouch~\cite{Lellouch:1995yv}
obtained stronger form factor bounds with statistical
analysis. We improved the bound using also the experimental $q^2$
spectrum from CLEO as additional inputs ~\cite{Fukunaga:2004zz}. 
After the BABAR measurement of the $q^2$ spectrum of $B\rightarrow\pi
l\nu$ decay~\cite{Athar:2003yg}, Arnesen et al. ~\cite{Arnesen:2005ez}
set $g_n(z) = z^n$ to obtain a simple bound on the
coefficients of the z-polynomial parameterization, which was further
improved imposing HQET power counting by Becher and
Hill~\cite{Becher:2005bg},~\cite{Hill:2006ub}. 

Since the BABAR measurement of the $q^2$ spectrum allows for the 
form factor shape determination, we also updated our 
determination of the $|V_{ub}|$ using the dispersive bound~\cite{Disp}. 
Using the form factor from HPQCD collaboration~\cite{Dalgic:2006dt} 
and the CLEO data, we obtain our preliminary estimate 
\begin{equation}
|V_{ub}|=\left[3.4^{+0.4}_{-0.6}\right]\times 10^{-3}
\end{equation}
\section{Summary}
%
%There have been major progresses in the unquenched QCD simulation in
%chiral regime, the renormalization schemes which allows for the
%nonperturbative determinations of the renormalization factors, and the
%new approach to the heavy quarks on the lattice. These developments 
%have been tested in light hadron physics or in quenched QCD and 
%are promising for improving the lattice calculation for B physics in
%near future.
There has been major progress in the unquenched
QCD simulation in the chiral regime, the renormalization
schemes which allows for nonperturbative
determinations of the renormalization factors,
and in the new approach to the heavy quarks on the lattice.
These developments have been tested in light
hadron physics or in quenched QCD and are promising
for improving the lattice calculation for B physics
in the near future.

%I reviewed recent results with $n_f=2$ dynamical overlap fermion by JLQCD
%collaboration. It was found that the chiral behavior of $m_{\pi}$ and
%$f_{\pi}$ are consistent the Next-to-next-leading order Chiral
%Perturbation theory. With the advent of the exact chiral symmetry 
%precise determination of $B_K$. 
I reviewed recent results with the $n_f$ = 2 dynamical
overlap fermion by the JLQCD collaboration. It was
found that the chiral behavior of $m_{\pi}$  and $f_{\pi}$ are consistent
with Next-to-next-leading order Chiral Perturbation
theory. With the advent of the exact chiral
symmetry a precise determination of BK was discussed.

%I also explained some new results in heavy quark physics such as $B^*
%B\pi$ coupling and the model independent determination of $|V_{ub}|$
%from dispersive bound. 
I also explained some new results in heavy quark
physics such as $B^* B\pi$ coupling and the model independent
determination of $|V{ub}|$ from the dispersive bound.

\begin{acknowledgments}
I would like to thank my colleagues in JLQCD collaboration.
Work is supported in part by the Grant-in-Aid of the
Ministry of Education (Nos. 19540286, 20039005).

\end{acknowledgments}

\end{document}